\documentclass[prd,amsmath,amssymb,groupedaddress,letterpaper,11pt,nofootinbib]{revtex4}
\usepackage{amsmath,amssymb,graphicx,bm,subfigure}

\begin{document}

\title{On the volatility of volatility}

\author{Stephen~D.~H.~Hsu} \email{hsu@duende.uoregon.edu}
\author{Brian~M.~Murray} \email{bmurray1@uoregon.edu}

\affiliation{Institute of Theoretical Science, University of Oregon,
  Eugene OR 94703-5203}

\begin{abstract}

The Chicago Board Options Exchange (CBOE) Volatility Index, VIX, is
calculated based on prices of out-of-the-money put and call options on
the S\&P 500 index (SPX). Sometimes called the ``investor fear
gauge,'' the VIX is a measure of the implied volatility of the SPX,
and is observed to be correlated with the 30-day realized volatility
of the SPX. Changes in the VIX are observed to be negatively
correlated with changes in the SPX.  However, no significant
correlation between changes in the VIX and changes in the 30-day
realized volatility of the SPX are observed. We investigate whether
this indicates a mispricing of options following large VIX moves, and
examine the relation to excess returns from variance swaps.

\end{abstract}

\maketitle

\section{Introduction} \label{intro}

Volatility is a fundamental characteristic of financial
markets. Although a derived quantity, describing the propensity of
prices to fluctuate, it plays an important role in options pricing and
in any simple characterization of market dynamics. In
Ref.~\cite{demeterfi} Demeterfi, et al. list three reasons for trading
volatility.  The first two involve direct speculation on the future
level of stock or index volatility. First, one may, due to a
particular directional view, simply want to be long or short
volatility. Second, one may want to speculate on the spread between
realized and implied volatility. Third, one may want to be long
volatility as a hedge against other portfolio components which are
effectively short volatility. For example, equity fund investors
following active benchmarking strategies, portfolio managers who are
judged against a benchmark, and risk arbitrageurs are all implicitly
short volatility. Due to their various circumstances, every one of
these types of market participants could stand to benefit, if they
could somehow add to their portfolios a long position on volatility.

Volatility swaps provide just such an opportunity. There is no cost to
enter these contracts. The payoff on the long side is equal to the
realized (annualized) volatility over the life of the contract minus a
fixed annualized volatility (the delivery or strike price) times a
notional amount of the swap in dollars per annualized volatility
point. Due to the square root relationship between volatility and
variance, and the more fundamental theoretical significance of
variance, it turns out to be easier to effectively price and hedge
variance swaps than volatility swaps. Therefore, we will primarily
focus our attention on variance swaps.

In order to replicate a variance swap, one needs to hold a portfolio
consisting of a particular distribution of options on the underlying
\cite{demeterfi}. On September 22, 2003, the Chicago Board Options
Exchange (CBOE) introduced the new CBOE Volatility Index (VIX). The
new VIX replaced an older volatility index that had a problematic
definition, and which will not be discussed further here. The new VIX
calculation is based on the prices of a batch of out-of-the-money and
near-the-money put and call options on the S\&P 500 index (SPX).
Indeed, the VIX has a very concrete economic meaning: it is the simply
the price of a linear portfolio of options. The square of the
VIX is the variance swap rate up to corrections due to the fact that
there are only SPX options at a finite number of strikes, as well as
the fact that there are occasional jumps in the underlying (SPX).  Put
another way, the square of the VIX is approximately equal to the
risk-neutral expectation of the annualized return variance over the
next 30 days, up to the corrections mentioned above \cite{carr}.
Interestingly, Carr and Wu \cite{carr} show that adding information
from a GARCH process to the information contained in the VIX does not
lead to a better prediction of the return variance than using the VIX
alone.

Therefore, it is logical to define 30-day variance swaps on the SPX as
contracts that depend on the difference between the realized variance
and the square of the VIX. It is well known that implied volatilities
are typically larger than realized volatilities\footnote{Implied
volatility which is systematically larger than realized volatility
would seem to provide a risk-free arbitrage, since it means all
options contracts are overpriced. In an idealized world of log-normal
price fluctuations, a trader could sell options contracts and hedge
away the risk by holding cash and the underlying. However, in the real
world, where volatility is itself volatile, there is no foolproof way
to completely hedge away the risk of selling an option. An option
seller is paid a premium to bear this risk, namely the systematic
difference between implied and realized volatilities.}. As the VIX is
an implied volatility of sorts (again, it is calculated based on
option prices; the Black-Scholes equation \cite{black_scholes} is not
used), one might guess that shorting variance swaps on the SPX may be
a successful investment strategy. Indeed, Carr and Wu \cite{carr}
verify that this is the case. But who would want to be on the long
side of these contracts? As mentioned above, there are a number of
market participants who are implicitly short volatility. Perhaps the
most common example is an equity fund. Over short timescales index
levels and implied volatility are often negatively correlated (this is
not the case over long timescales, however). An investor whose
portfolio consists primarily of equities is willing to pay a premium
to be on the long side of a variance swap. This is analogous to
insurance, where the party on the long side of the contract is happy
to pay a relatively small premium over an extended period of time,
with the assurance that if something goes wrong (i.e., the SPX crashes
in the case of a variance swap, one's house burns down in the case of
an insurance policy), they will be compensated for their loss because
they are long variance (or long an insurance policy).

It is rather striking that there is often a negative correlation between
changes in implied volatility and changes in the underlying. This
implies a non-trivial memory in price dynamics which goes beyond the
most naive (i.e., log-normal) models. Black \cite{black} first
proposed the ``leverage effect'' as a possible explanation of this
negative correlation.  The idea is essentially that at a fixed level
of debt, a decline in equity level increases the leverage and
therefore the risk for which implied volatility is a proxy.  However,
this mechanism is probably too small to explain the entire effect, at
least in the case of the SPX, where a one percent change in the index
causes a roughly negative four percent change in the implied
volatility. Other possible explanations include put gouging (when the
market has moved down, and demand for insurance is high) and call
overwriting (when markets are up).

The plan of this article is as follows. In Sec.~\ref{imp_real}
correlations between the SPX, the VIX, and the 30-day realized
volatility are examined, as are correlations between changes in these
quantities. We note that there is no significant correlation between
changes in the VIX and changes in the 30-day realized volatility.
This suggests that, at least in theory, some options are mispriced
after large moves in the index. In Sec.~\ref{vol_surf} we briefly
discus the behavior of the volatility surface on days when there is a
large change in the VIX and the SPX. In Sec.~\ref{trading} we examine
shorting variance swaps on the SPX as in Carr and Wu \cite{carr}. We
investigate a trading strategy in which a large change in the VIX is
used as a signal for selectively shorting variance swaps. The lack of
correlation identified in Sec.~\ref{imp_real} between changes in the
VIX and changes in the 30-day realized volatility suggests that this
strategy would outperform simply continuously shorting variance
swaps. However, because of the large premium (excess return)
associated with variance swaps, the additional advantage is relatively
small. Finally, we summarize our results in Sec.~\ref{conclusion}.

\begin{figure}
\includegraphics[width=14cm]{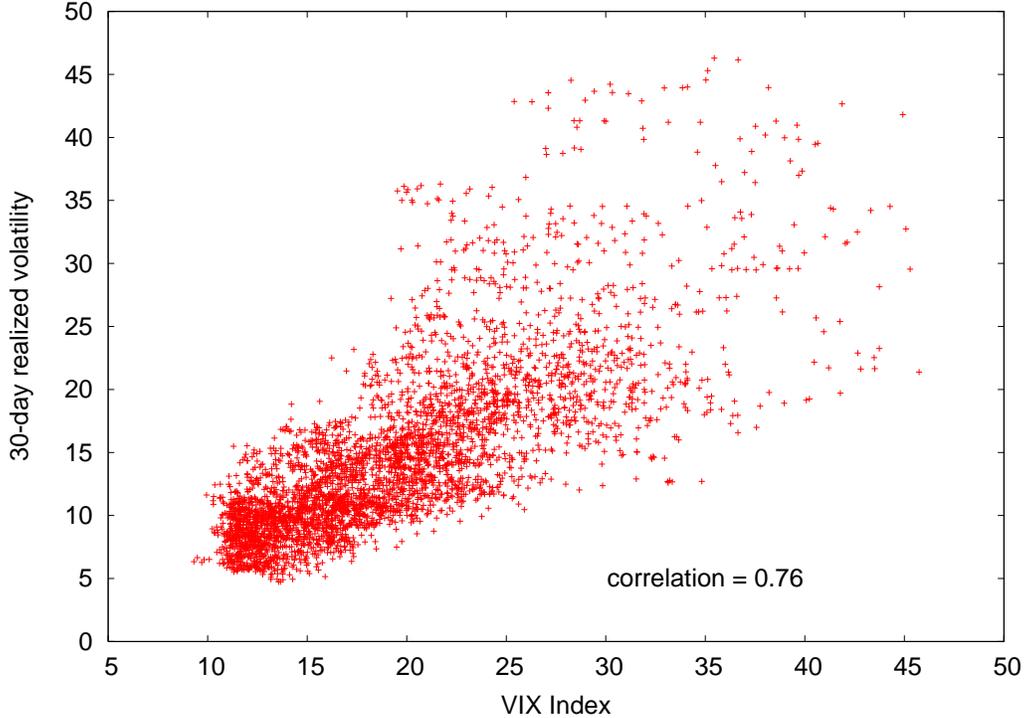}
\caption{The 30-day realized volatility of the SPX versus the VIX for
2 Jan 1990 to 29 Jun 2006. A significant level of correlation is
observed between the realized and implied volatilities.}
\label{vix_rv9006_cor}
\end{figure}

\begin{figure}
\includegraphics[width=14cm]{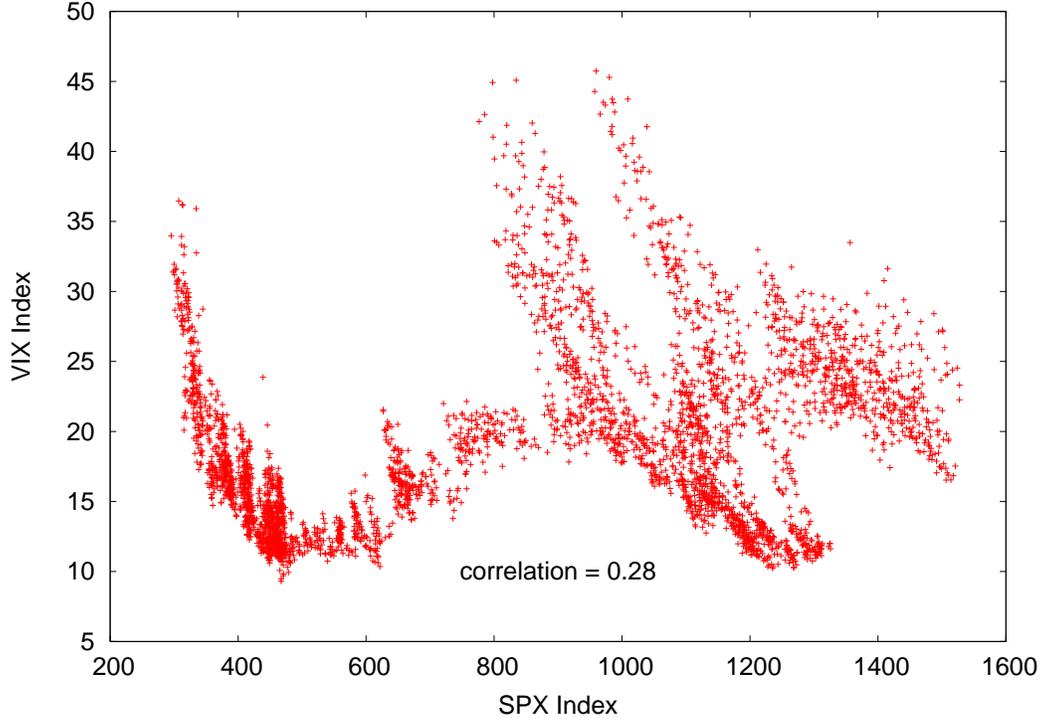}
\caption{The VIX versus the SPX for 2 Jan 1990 to 29 Jun 2006. Over
this long timescale, no significant correlation between the indexes is
observed.}
\label{gspc_vix9006_cor}
\end{figure}

\begin{figure}
\includegraphics[width=14cm]{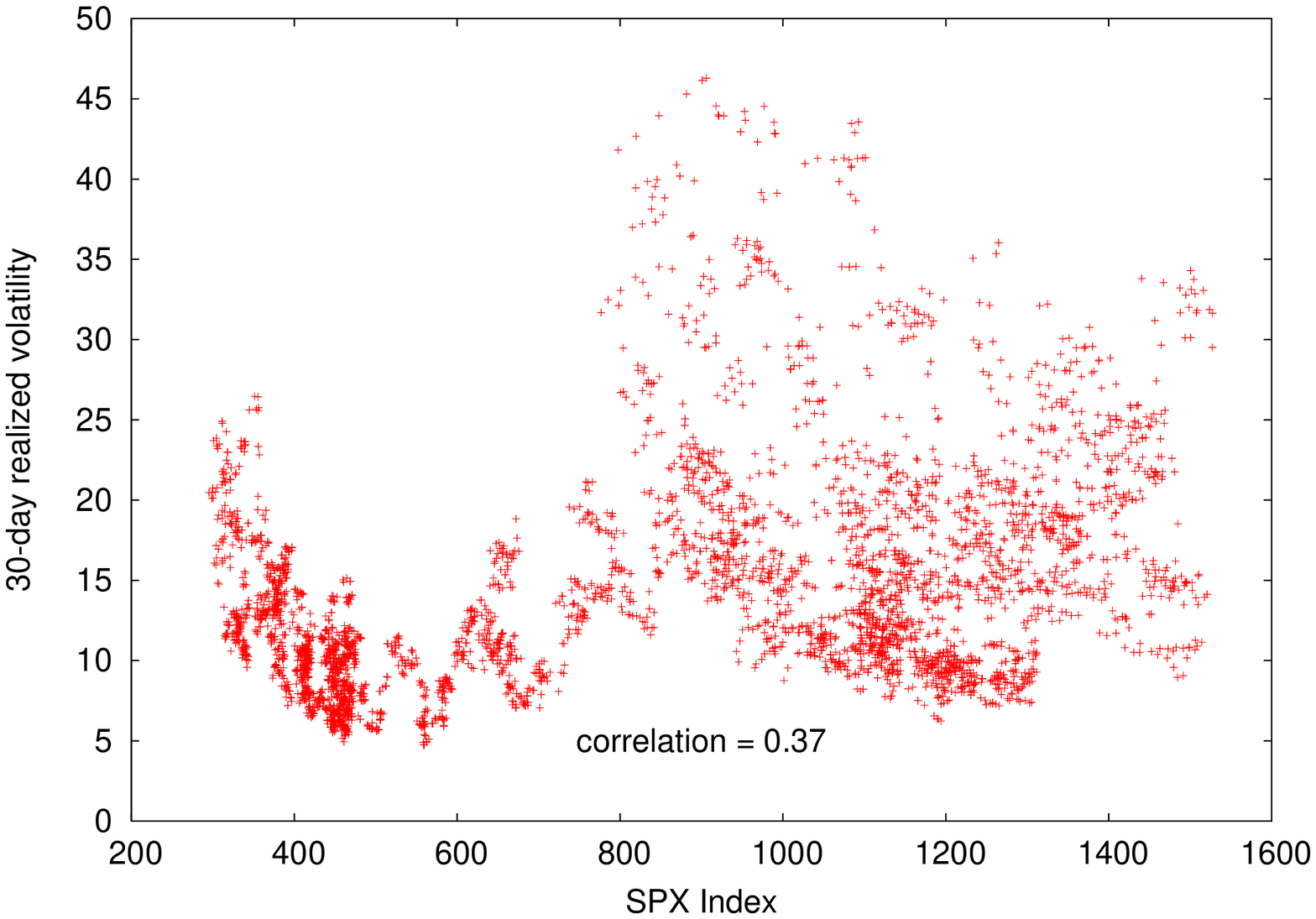}
\caption{The 30-day realized volatility of the SPX versus the SPX for
2 Jan 1990 to 29 Jun 2006. Over this long timescale, no significant
correlation is observed.}
\label{gspc_rv9006_cor}
\end{figure}

\section{Implied, realized, and the underlying}\label{imp_real}

The sample period for calculating the 30-day realized volatility
RVol$_{t,t+30}$ (defined below) consists of 4178 trading days from 2
Jan 1990 to 31 Jul 2006. For the SPX and the VIX, we use 4157 trading
days, from 2 Jan 1990 to 29 Jun 2006. SPX and VIX data were obtained
from Yahoo Finance.

\subsection{The SPX, the VIX, and realized volatility}\label{imp_real_a}

We define the 30-day realized volatility as follows:
\begin{eqnarray}
  {\rm RVol}_{t,t+30} = 100\times\sqrt{\frac{365}{30}\sum_{i=1}^{30}
                 \left[\ln\left(\frac{S_{t+i}}{S_{t+i-1}}\right)\right]^2}.
\end{eqnarray}
As is common practice, we use a definition which assumes zero mean.

We use the standard definition of the correlation between two time series
$X_t$ and $Y_t$ consisting of $n$ points $x_i$ and $y_i$, respectively:
\begin{eqnarray}
  {\rm Cor}\left(X_t,Y_t\right)= 
    \frac{\sum_{i=1}^n (x_i-\bar{x})(y_i-\bar{y})}{(n-1) s_x s_y},
\end{eqnarray}
where $\bar{x}$ and $\bar{y}$ are the usual sample means, and $s_x$
and $s_y$ are the usual sample standard deviations.

We find a significant correlation between the VIX and the 30-day realized
volatility:
\begin{eqnarray}
  {\rm Cor}\left({\rm VIX}_t,{\rm RVol}_{t,t+30}\right) = 0.76.
\end{eqnarray}

Scatter plots of combinations of the VIX, the 30-day realized
volatility, and the SPX are shown in Figs.~\ref{vix_rv9006_cor},
\ref{gspc_vix9006_cor}, and \ref{gspc_rv9006_cor}.  The correlations
between the SPX and both realized and implied volatilities are not
significant over long timescale of our sample (16 and a half
years).

\subsection{Changes in the SPX, the VIX, and realized volatility}\label{imp_real_b}

In order to examine correlations between changes in the indexes, we
make the following definitions:
\begin{eqnarray}
  {\rm CSPX}_t = \frac{{\rm SPX}_t - {\rm SPX}_{t-1}}{{\rm SPX}_{t-1}}
           \nonumber \\
  {\rm CVIX}_t = \frac{{\rm VIX}_t - {\rm VIX}_{t-1}}{{\rm VIX}_{t-1}}.
\end{eqnarray}

\begin{figure}
\includegraphics[width=14cm]{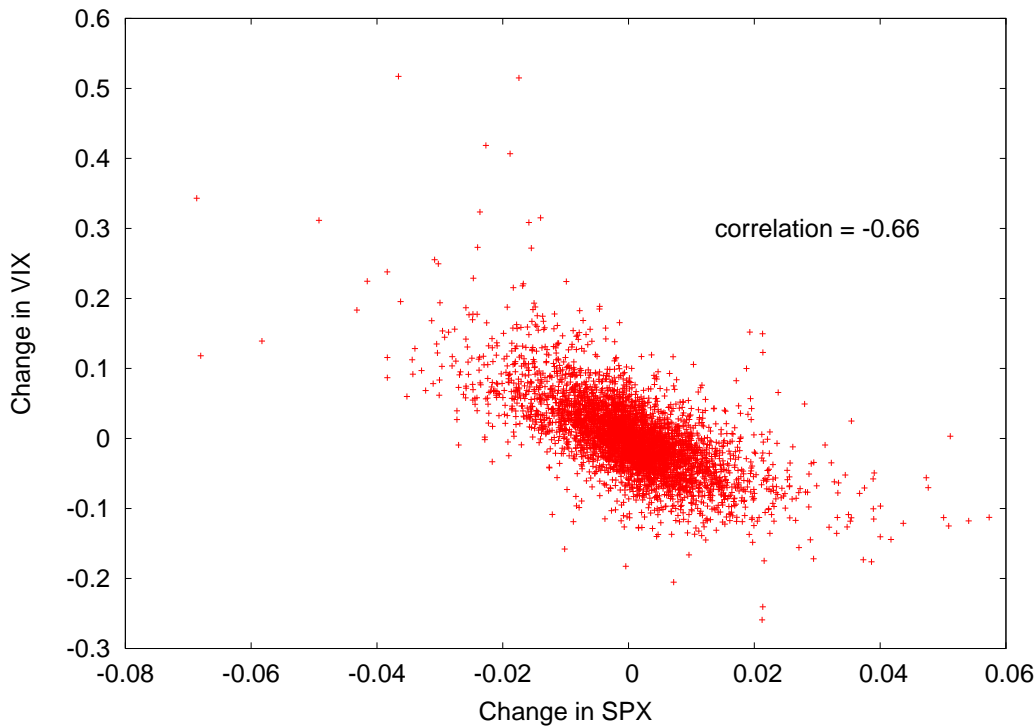}
\caption{Change in the VIX versus change in the SPX for 2 Jan 1990 to
29 Jun 2006. Changes in the indexes have a significant level of
negative correlation.}
\label{gspc_vix9006_comb}
\end{figure}

There is a significant negative correlation between changes in the
VIX and changes in the SPX:
\begin{eqnarray}
  {\rm Cor}\left({\rm CVIX}_t,{\rm CSPX}_t\right) = -0.66.
\end{eqnarray}
See Fig.~\ref{gspc_vix9006_comb} for a scatter plot of CVIX$_t$ versus
CSPX$_t$. Note that this negative correlation is interesting, as it
implies that there is a non-trivial memory in price dynamics which
goes beyond the most naive (i.e., log-normal) models.

\begin{figure}
\includegraphics[width=14cm]{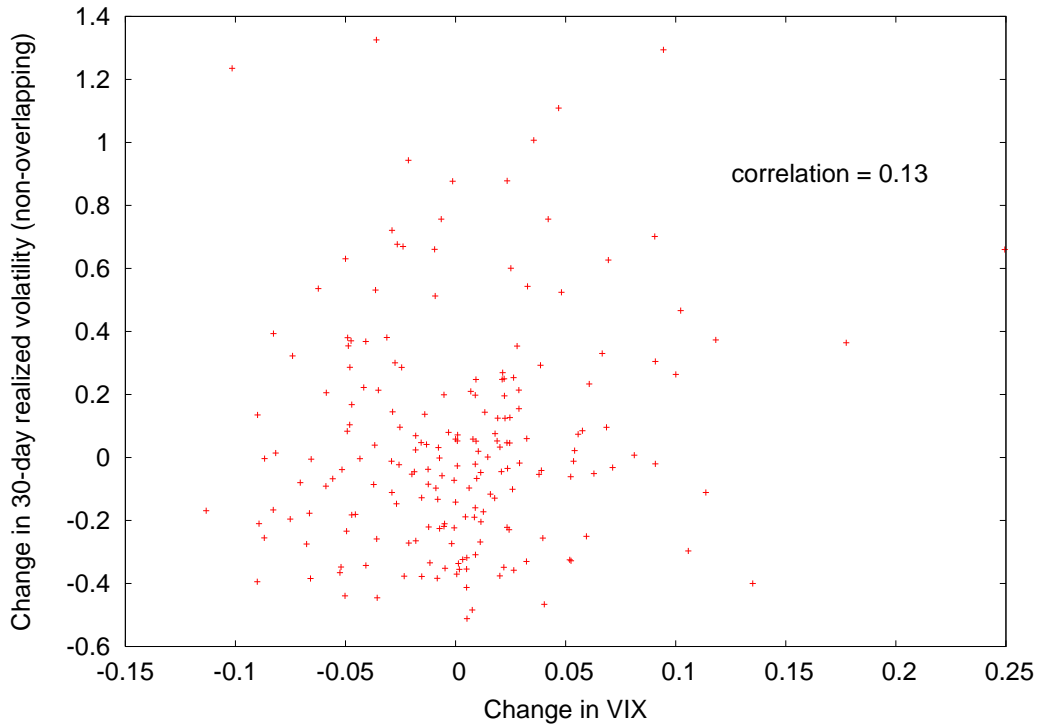}
\caption{Change in the 30-day realized volatility versus change in the
VIX for non-overlapping intervals from 1 Feb 1990 to 29 Jun 2006. With
a correlation of 0.13, this dataset ($s=2$, see text) provides a
typical example of the lack of correlation between changes in the VIX
and changes in the 30-day realized volatility. See
Fig.~\ref{cvix_crv9006_noover_cor_hist} for a histogram of the
correlations for all possible choices of the offset $s$.}
\label{cvix_crv9006_noover}
\end{figure}

In order to examine the correlation between changes in the VIX and
changes in the 30-day realized volatility, we must first find a
suitable definition for the change in the 30-day realized volatility.
To this end, we define:
\begin{eqnarray}
  {\rm CRVol}_{t} = \frac{{\rm RVol}_{t,t+30} - {\rm RVol}_{t-31,t-1}}
                             {{\rm RVol}_{t-31,t-1}}.
\end{eqnarray}
This quantity compares the 30-day realized volatility of the 30-day
time period ending at time to time $t-1$ with that of the 30-day time
period beginning at time $t$. It is an appropriate way of measuring
the change in the 30-day realized volatility because it compares
volatilities of two {\it independent} neighboring time
periods. However, it remains true that CRVol$_t$ and CRVol$_{t+1}$ are
not independent. Therefore, in estimating the correlation between the
change in the VIX and the change in the 30-day realized volatility, it
would not be appropriate to simply compare the two time series
CRVol$_t$ and CVIX$_t$. Instead, we compare the correlation between
CRVol$_{s+30t}$ and CVIX$_{s+30t}$ for all offsets $s$ with
$0<s<30$. All 21 such time series consist of 196 days (there are 21
trading days for every 30 calendar days). See
Fig.~\ref{cvix_crv9006_noover} for a scatter plot of CRVol$_{s+30t}$
versus CVIX$_{s+30t}$ for the case $s=2$, i.e., the first change in
the 30-day realized volatility considered is between the 30 days prior
to and the 30 days following 5 Feb 1990. For this choice of offset,
the correlation is typical of the values obtained for all values of
$s$, with a value of 0.13.

\newpage

\begin{figure}
\includegraphics[width=14cm]{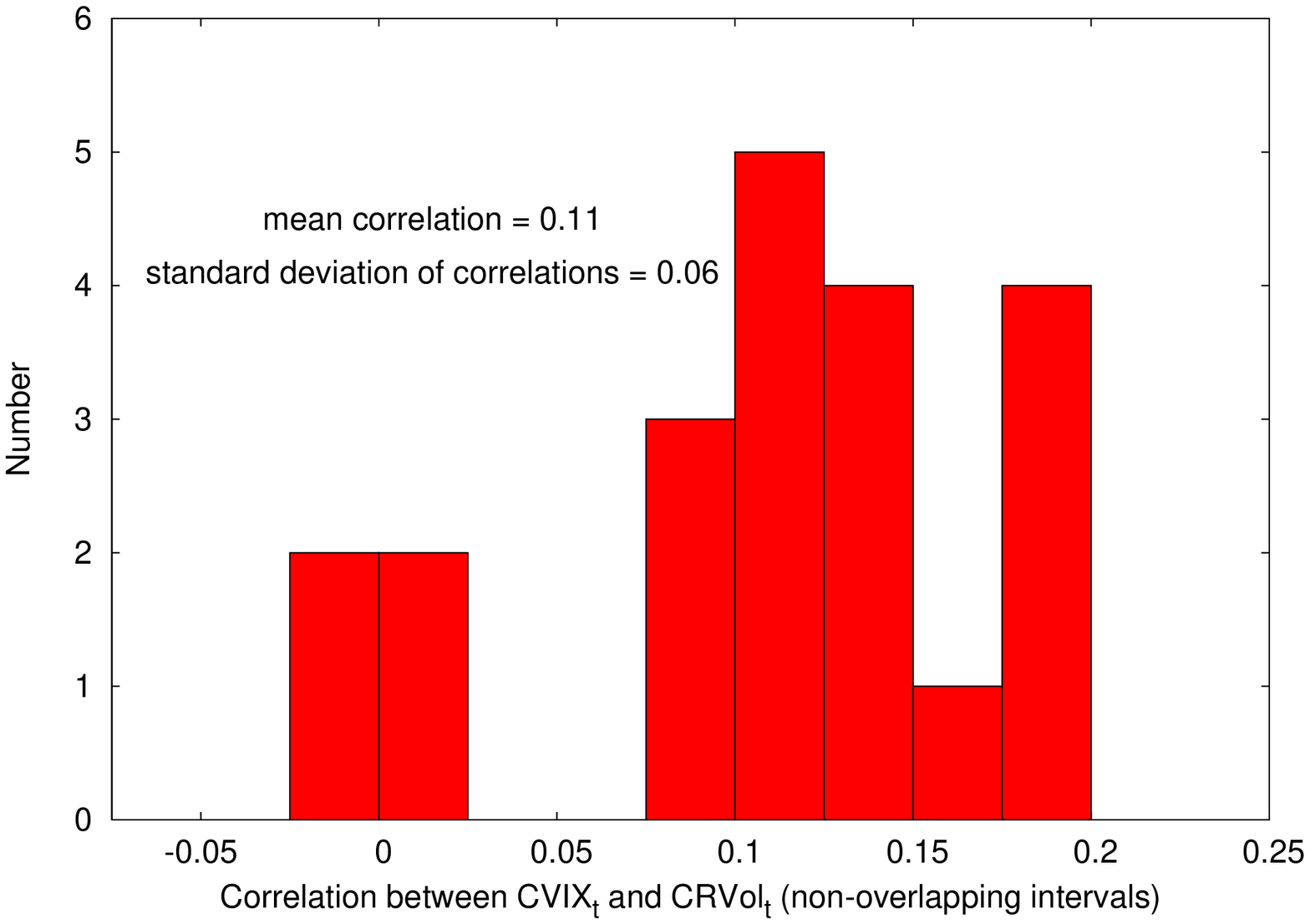}
\caption{Correlations between changes in the VIX changes in the 30-day
realized volatility computed for 21 different non-overlapping
intervals, each one consisting of 196 entries.}
\label{cvix_crv9006_noover_cor_hist}
\end{figure}

A histogram of the correlations for all possible values of offset $s$
is shown in Fig.~\ref{cvix_crv9006_noover_cor_hist}. Each of the 21
correlations computed is correlating two time series (CVIX$_t$ and
CRVol$_t$) of 196 days each. With a mean correlation of 0.11 and a
standard deviation of the correlations of 0.06, it is clear that {\it
a change in the VIX does not predict a change in the 30-day realized
volatility of the SPX.}

\section{The behavior of implied volatility}\label{vol_surf}

In order to try to determine the cause of the negative correlation
between changes in the SPX and changes in the VIX, we obtained closing
prices for near-the-money SPX put and call options on the days
immediately before and the days of the four largest VIX increases
and four largest VIX decreases of 2005 and 2006. The data were
obtained from Bloomberg.

\begin{figure}[ht]
\includegraphics[width=14cm]{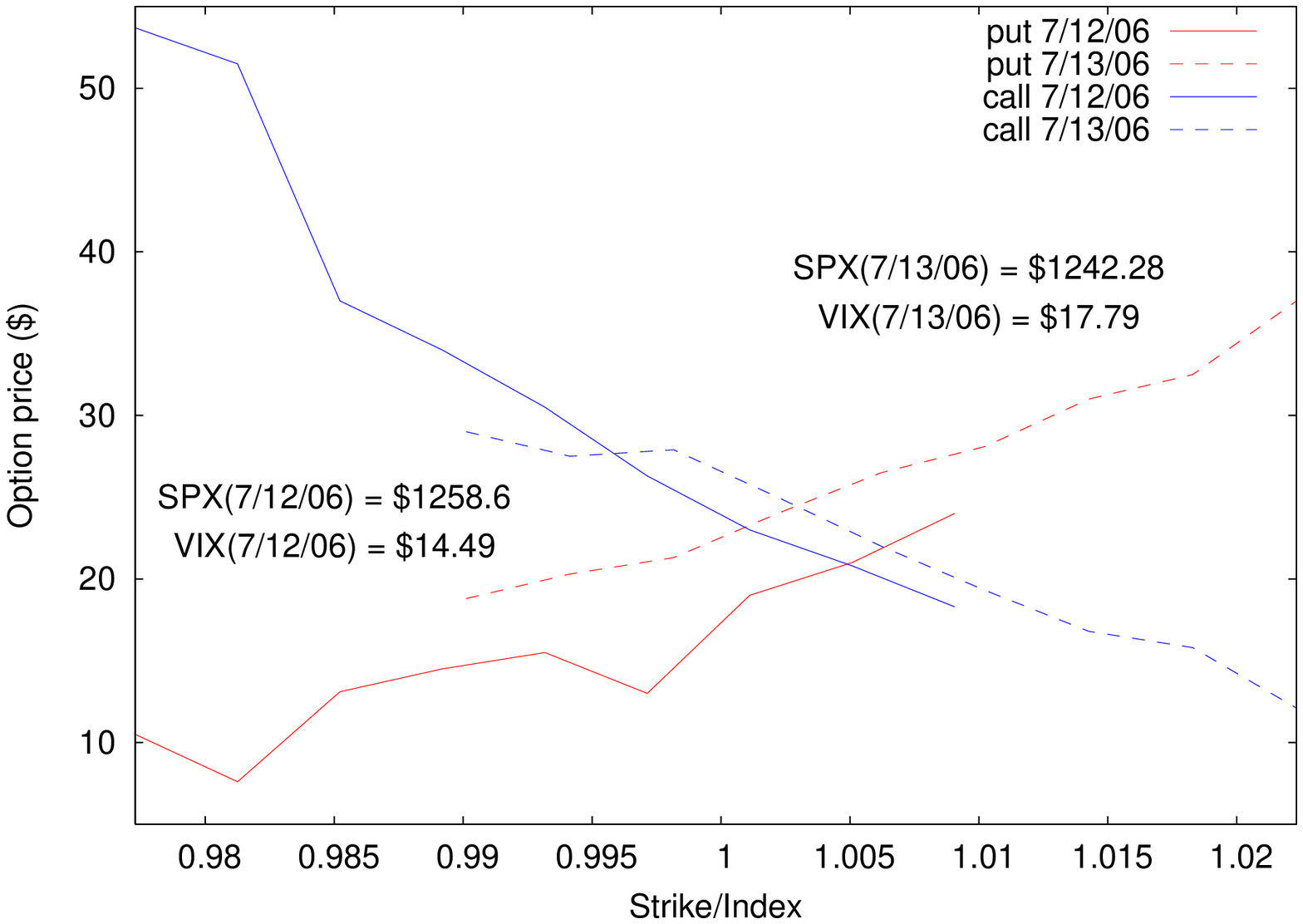}
\caption{Closing prices for put and call options on the SPX on
the day immediately before and the day of a large VIX increase.}
\label{0806_71206}
\end{figure}

\begin{figure}[ht]
\includegraphics[width=14cm]{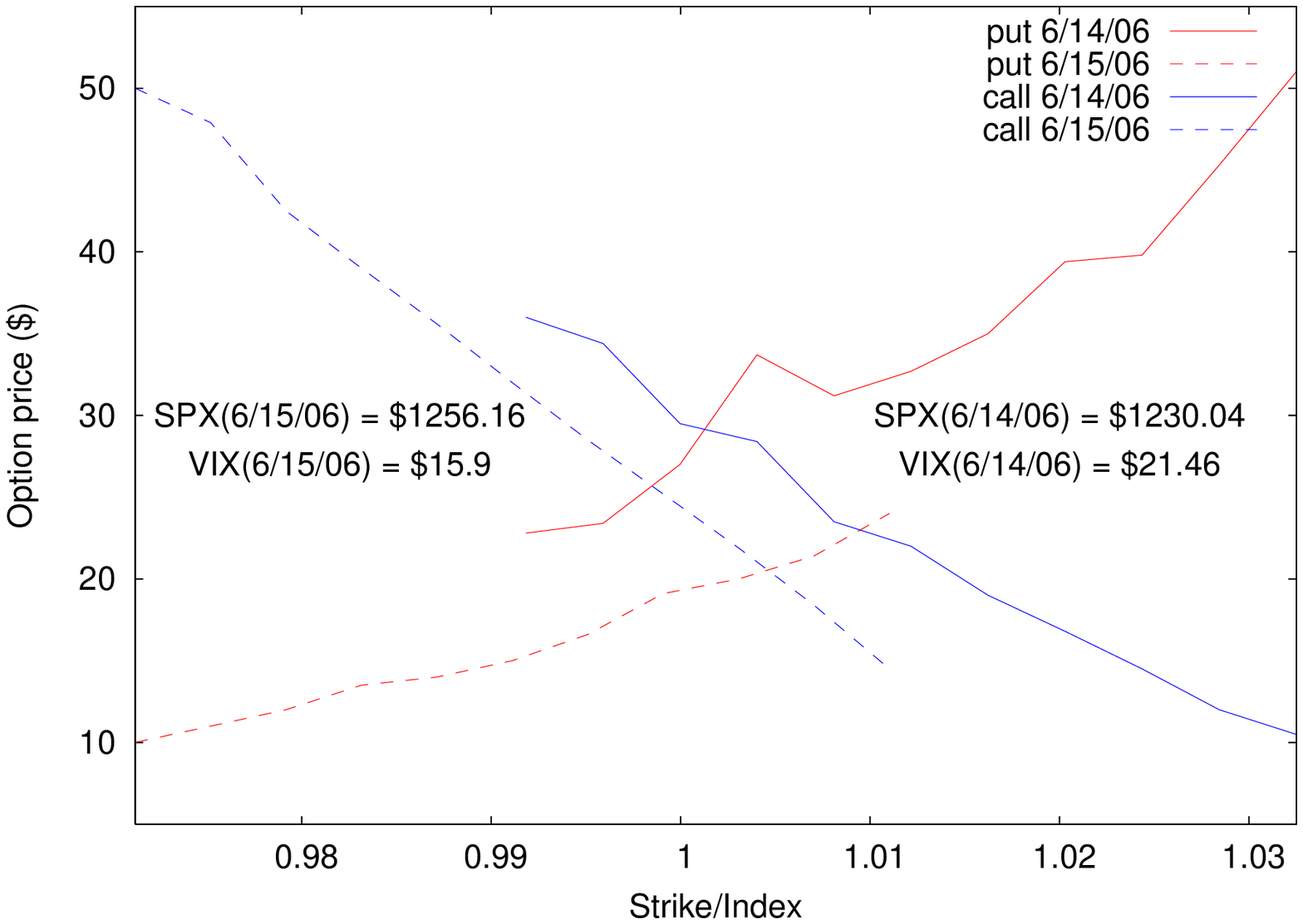}
\caption{Closing prices for put and call options on the SPX on
the day immediately before and the day of a large VIX decrease.}
\label{0706_61406}
\end{figure}

\begin{figure}[ht]
\includegraphics[width=14cm]{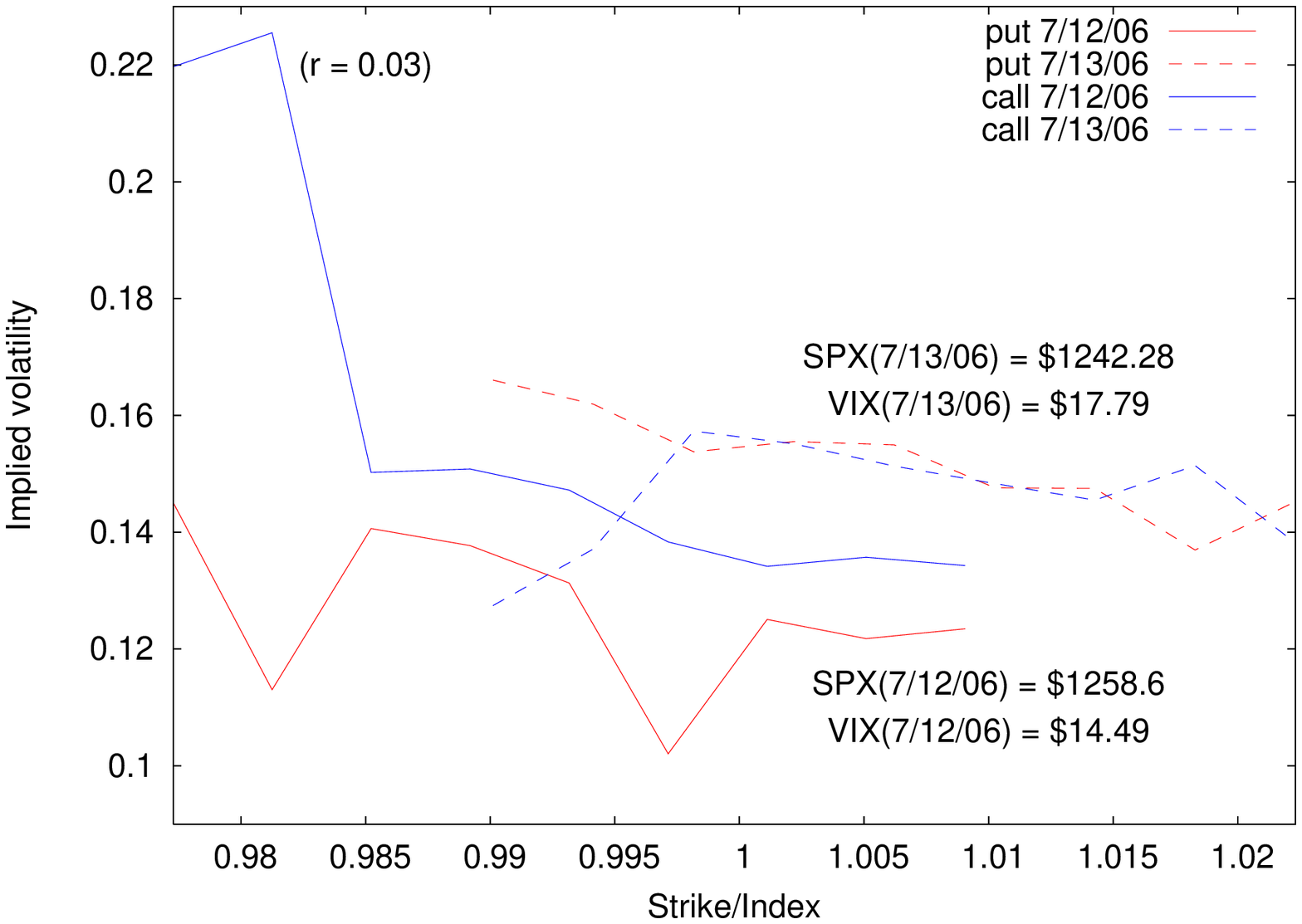}
\caption{Black-Scholes implied volatility for put and call options,
the day before and the day of a large VIX increase. A volatility skew
is observed, with implied volatility decreasing with strike price for
both puts and calls, both before and after the big change. An
annualized risk-free rate of 0.03 is assumed.}
\label{0806_71206vs_r3_n}
\end{figure}

\begin{figure}[ht]
\includegraphics[width=14cm]{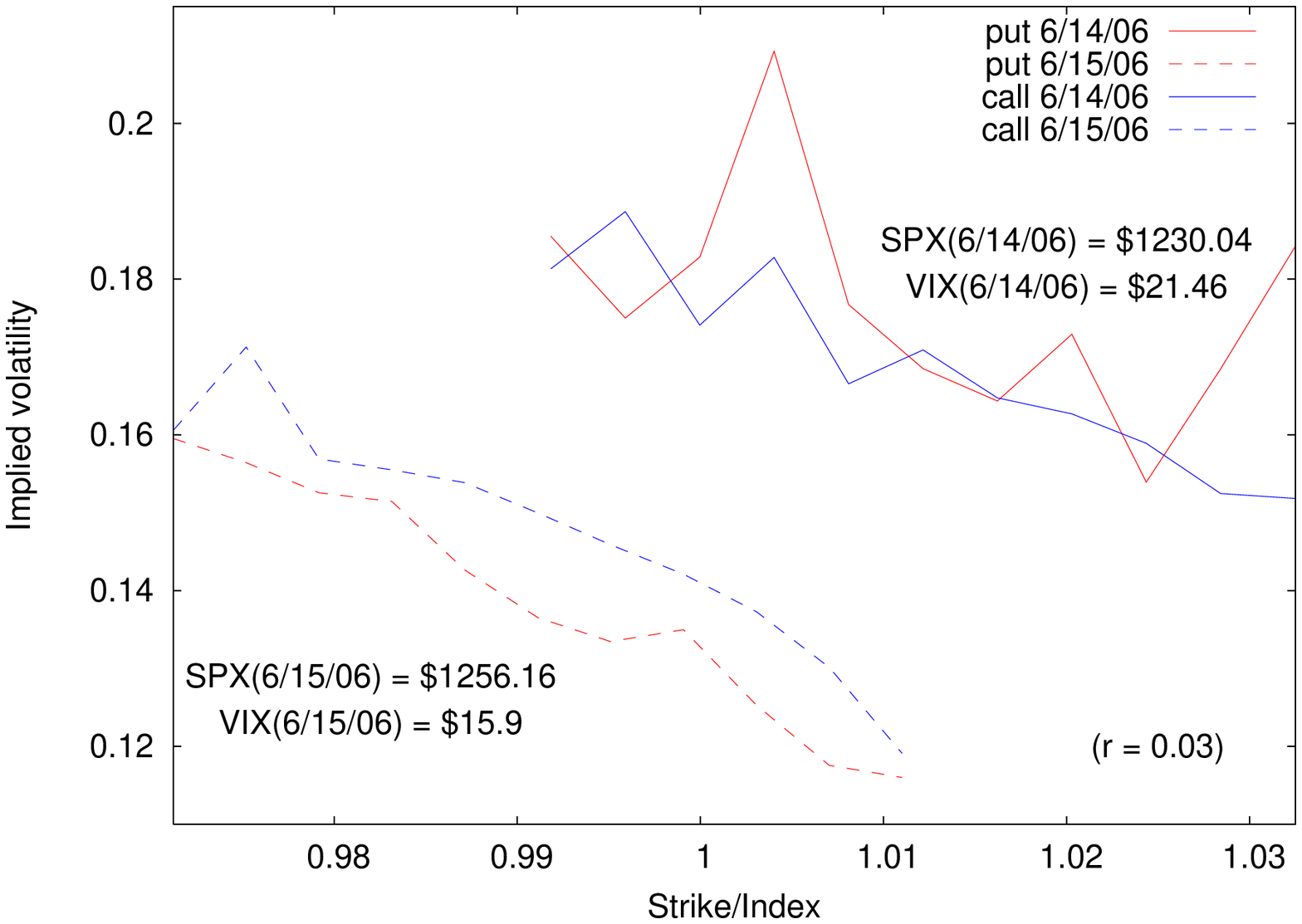}
\caption{Black-Scholes implied volatility for put and call options,
the day before and and the day of a large VIX decrease. A volatility
skew is observed, with implied volatility decreasing with strike price
for both puts and calls, both before and after the big move. An
annualized risk-free rate of 0.03 is assumed.}
\label{0706_61406vs_r3_n}
\end{figure}

In Figs.~\ref{0806_71206} and \ref{0706_61406} we plot option price
versus the ratio of strike price to index level for both puts and
calls the day before and the day of one of the largest VIX increases
and decreases, respectively, of 2006. As expected, both puts and calls
at a given distance from at-the-money become more expensive when VIX
increases and less expensive when VIX decreases.

\newpage

Similarly, the volatility surface is plotted in
Figs.~\ref{0806_71206vs_r3_n} and \ref{0706_61406vs_r3_n}. Although
we see that, as expected, Black-Scholes implied volatility increases
or decreases when the VIX does, it is difficult to comment as to
{\it why} the VIX is changing, i.e., which particular options are
causing the VIX to increase or decrease. Also, we see roughly
linear volatility skews, as have been present in many indexes, since
the 1987 crash.

Without much more data, we cannot say which options cause the VIX to
change. Put gouging, call overwriting or Black's ``leverage effect''
may be at work here, but we cannot say with any certainty. It would be
interesting to further investigate the volatility surface.

\section{Trading realized and implied volatility}\label{trading}

\begin{figure}
\includegraphics[width=14cm]{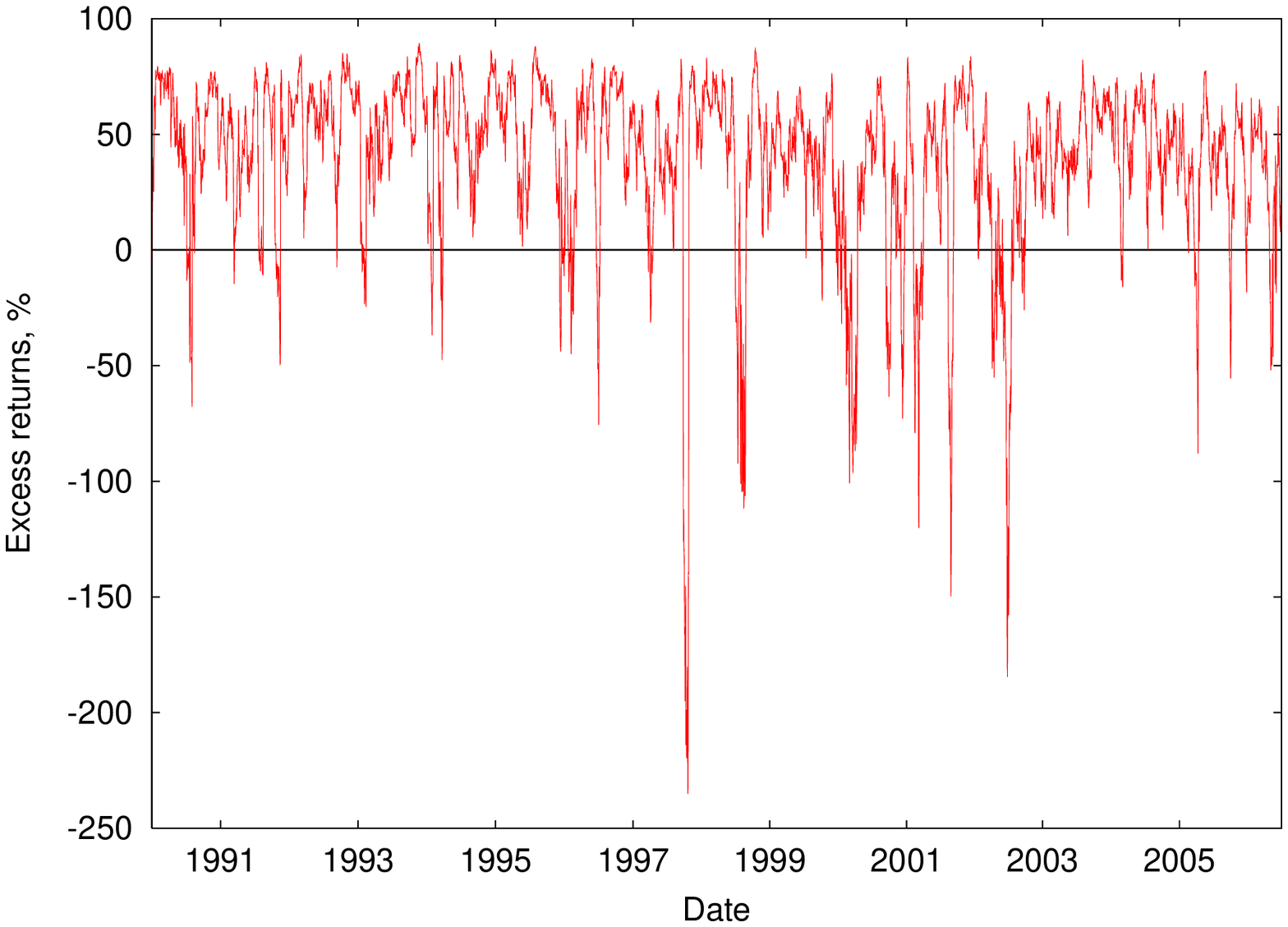}
\caption{Excess returns from shorting 30-day variance swaps as in Carr
and Wu \cite{carr}.}
\label{vix_rv9006_er}
\end{figure}

\begin{figure}
\includegraphics[width=14cm]{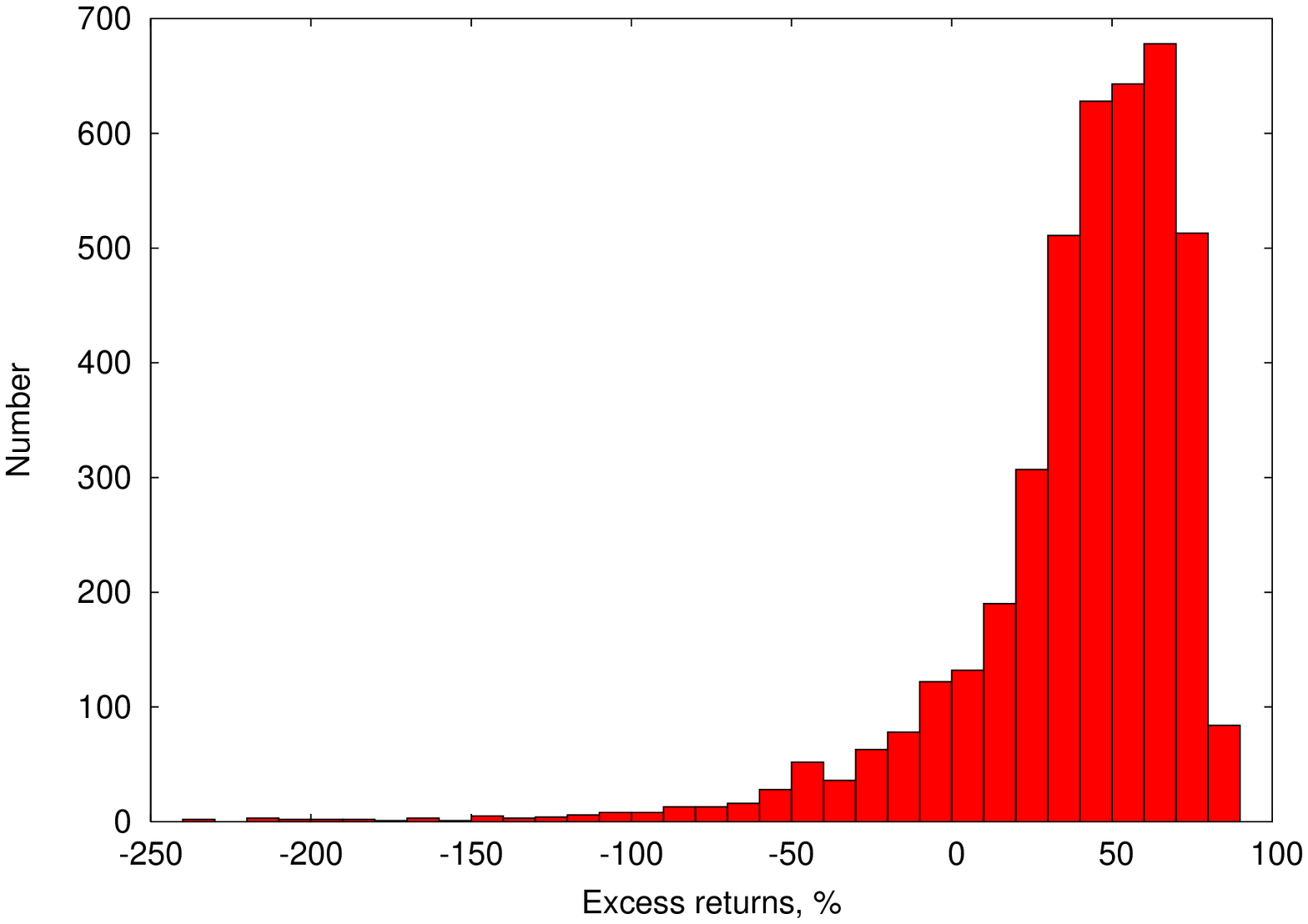}
\caption{Histogram of excess returns from shorting 30-day variance
swaps as in Carr and Wu \cite{carr}. The mean excess return is
estimated at 39.45 percent, with a standard deviation of 36.62
percent.}
\label{vix_rv9006_er_hist}
\end{figure}

As stated in Sec.~\ref{intro}, the VIX has a concrete economic
meaning: its square is the variance swap rate up to corrections due to
the fact that there are only SPX options at a finite number of
strikes, as well as the fact that there are occasional jumps in the
underlying (the SPX). In Ref.~\cite{carr}, Carr and Wu use this
fact to determine the excess returns that would have been gained
from shorting variance swaps on every day of the sample period, where
the excess return is defined as:
\begin{eqnarray}
  {\rm ER}_{t,t+30} = 100\times
    \frac{{\rm VIX}_t^2 - {\rm RVol}_{t,t+30}^2}{{\rm VIX}_t^2}.
\end{eqnarray}
See Figs.~\ref{vix_rv9006_er} and \ref{vix_rv9006_er_hist} for a
time series and a histogram, respectively, of the excess returns.

It is worth emphasizing the fact that the mean excess return gained
from continuously shorting variance swaps is large, at nearly 40
percent.  Why is this premium so large?  In order to
answer this question, one must take into account that the distribution
of excess returns is heavily skewed; there are a number of occurrences
of very large negative excess return. As discussed in
Sec.~\ref{intro}, parties on the long side of variance swaps are
willing to pay a high premium for the insurance provided against
periods of high realized volatility (relative to the VIX). Whether
they are portfolio managers who are judged against a benchmark, equity
funds or others, they are effectively short volatility, and are
therefore willing to pay for the insurance that variance swaps
provide. Interestingly, Carr and Wu \cite{carr} argue that the CAPM
cannot fully account for the size of the excess return associated with
variance swaps. This perhaps indicates an inefficiency
or mispricing in this market.

\newpage

In Sec.~\ref{imp_real_b} it was shown that there is no correlation
between changes in the VIX and changes in the 30-day realized
volatility of the SPX. This means that the excess returns gained from
shorting 30-day variance swaps would increase if the swaps are shorted
only on days when there is a large increase in the VIX as opposed to
every day as in Carr and Wu \cite{carr}. This is true because a change
in the VIX does not predict a change in the 30-day realized
volatility. Therefore, on average the payoff from shorting the swap
will be higher when the VIX has recently increased. The opposite
should be true as well: shorting swaps only on days when there is a
large decrease in the VIX should lead to a decrease in the excess
returns.

\begin{table}
\begin{center}
\begin{tabular}{|c|c|c|c|}
 \hline  
 Number of days swaps are shorted & $\overline{ER}$ & $s_{ER}$ 
                       & $\overline{ER}/s_{ER}$ \\ \hline
 4157 (entire sample) & 39.45 & 36.62 & 1.08 \\ \hline
 415 (largest VIX increases) & 40.43 & 35.79 & 1.13 \\ \hline
 83 (largest VIX increases) & 43.20 & 35.03 & 1.23 \\ \hline
 415 (largest VIX decreases) & 35.25 & 40.81 & 0.86 \\ \hline
 83 (largest VIX decreases) & 31.10 & 46.53 & 0.67 \\ \hline
\end{tabular}
\end{center}
\caption{Average excess return $\overline{ER}$, the standard deviation
of the excess returns $s_{ER}$, and the ratio of the two for various
trading strategies. Strategies involving the shorting of 30-day
variance swaps only on days when there is a large increase in the VIX
slightly outperform a strategy in which the swaps are shorted on every
day. Shorting only on days when there is a large decrease in the VIX does
worse.}
\label{tab}
\end{table}

Table~\ref{tab} shows the average excess return $\overline{ER}$, the
standard deviation of the excess returns $s_{ER}$, and the ratio of
the two for various trading strategies. As predicted based on the
independence of changes in the VIX relative to changes in the 30-day
realized volatility, we see that strategies involving the shorting of
30-day variance swaps only on days when there is a large increase in the VIX
slightly outperform a strategy in which the swaps are shorted on
every day, while shorting only on days when the VIX experiences a large
decrease does worse.

The effect, however, appears to be small. Presumably, this is the case
because the average excess return is so large for the simplest trading
strategy in which the swaps are shorted every day. Even in the case
where one shorts only on the days with the largest one percent of VIX
increases, the relative improvement in the average excess return does
not appear to be significant. In addition, we considered the possibility
of further restricting the trading strategy such that one is only
engaged in one swap contract at a time. Again, such a strategy does
not do significantly better than the strategy of shorting every day.

\section{Conclusion} \label{conclusion}

We investigated a number of features of implied and realized
volatility of the SPX index. Sec.~\ref{imp_real} examines
correlations between the SPX, the VIX, and the 30-day realized
volatility of the SPX, as well as between changes in these
quantities. We confirmed that the VIX and the 30-day realized
volatility are correlated, and that while changes in the SPX are
negatively correlated with changes in the VIX, the levels of the two
indexes are not correlated over the nearly 16 years of data that we
analyzed (although they may be negatively correlated on shorter
timescales).  Interestingly, as shown in
Fig.~\ref{cvix_crv9006_noover_cor_hist}, we found no significant
correlation between changes in the VIX and changes in the 30-day
realized volatility of the SPX. This means that short term changes in
the VIX do not correctly predict the actual realized volatility, and
suggests that at least some options are mispriced after large moves in
the index.

The details of the negative correlation between changes in the SPX and
changes in the VIX were addressed in Sec.~\ref{vol_surf}. Without a
large dataset of historical options prices, it is difficult to
identify the cause of the negative correlation between changes in the
SPX and changes in the VIX. To this end, it would be interesting to
examine the volatility surface in more detail on days with large
market moves.

We returned to the issue of correlation between realized and implied
volatility in Sec.~\ref{trading}. We began by reproducing the analysis
of Carr and Wu \cite{carr} regarding the excess return obtained by
continuously shorting variance swaps on every trading day of the
sample period. We then analyzed whether improved returns could be
gained by selectively shorting variance swaps using large changes in
the VIX as a signal. In the insurance analogy, this strategy would be
similar to an insurance company carefully selecting when to sell
insurance policies based on their expectations about the excess
returns to be had given a particular trigger criterion (e.g., selling
hurricane insurance when demand is high, but the intrinsic probability
of a storm has not changed from its historical value). The lack of
correlation identified in Sec.~\ref{imp_real} between changes in the
VIX and changes in the 30-day realized volatility suggests that this
strategy would outperform simply continuously shorting variance
swaps. This appears to be the case although statistics are
limited. Due to the large premium (excess returns) already associated
with variance swaps, we find that the additional advantage is
relatively small.

\section{Acknowledgements}

We thank Myck Schwetz (PIMCO) for useful comments and some help with
historical data, and Thomas Gould (CSFB) for additional feedback.

\end{document}